\documentclass[letterpaper,12pt]{JHEP3}

\usepackage{graphicx}
\usepackage{amsmath}
\usepackage{amssymb}
\def\bZ{\mathbb{Z}}
\def\cN{\mathcal{N}}
\def\cA{\mathcal{A}}
\def\cB{\mathcal{B}}
\def\cC{\mathcal{C}}
\def\cD{\mathcal{D}}
\def\cO{\mathcal{O}}

\def\cD{\mathcal{D}}

\def\tr{\mathop{\mathrm{tr}}}
\def\Im{\mathop{\mathrm{Im}}}

\def\vev#1{\langle#1\rangle}
\def\half{\hbox{$\frac 12$}}
\def\quarter{\hbox{$\frac 14$}}
\def\SU{{\text{SU}}}
\def\USp{{\text{USp}}}
\def\UONE{{\text{U(1)}}}

\preprint{
\hbox{}\hfill arXiv:0804.1957}

\title{Central charges of $\cN=2$
superconformal field theories in four dimensions}

\author{
Alfred D. Shapere$^{1,2}$ and Yuji Tachikawa$^1$\\

\bigskip

$^1$ School of Natural Sciences, Institute for Advanced Study,\\
 Princeton,  New Jersey 08540, USA
 
\medskip

$^2$ Department of Physics and Astronomy, University of Kentucky, \\
Lexington, Kentucky 40506-0055, USA

}

\abstract{
We present a general method for computing the central charges $a$ and $c$ of $\cN=2$ superconformal field theories corresponding to singular points in the moduli space of $\cN=2$ gauge theories.  Our method relates $a$ and $c$ to the $\UONE_R$ anomalies of the topologically twisted gauge theory.  We evaluate these anomalies by studying the  holomorphic dependence of the path integral measure on the moduli. We calculate $a$ and $c$ for superconformal points in a variety of gauge theories, including $\cN=4$ $\SU(N)$, 
$\cN=2$ pure $\SU(N)$ Yang-Mills,
and $\USp(2N)$ with 1 massless antisymmetric and 4 massive fundamental hypermultiplets.
In the latter case, we reproduce the conformal and flavor central charges previously calculated using the gravity duals of these gauge theories.
For any SCFT in the class under consideration, we derive a previously conjectured expression for $2a-c$ in terms of the sum of the dimensions of operators parameterizing the Coulomb branch.  Finally, we prove that the ratio $a/c$ is bounded above by 5/4 and below by 1/2.
}


\begin{document}

\section{Introduction}

Seiberg and Witten's \cite{SW1,SW2} approach to solving
$\cN=2$ gauge theories in the Coulomb phase has led to
enormous advances in our understanding of strong-coupling dynamics in four dimensions.   In particular, we now know of many examples of strong-coupling fixed points with $\cN=2$ superconformal symmetry \cite{AD,APSW,EHIY,EH,MN1,MN2,ACSW,AWI}.
The dynamics of such a fixed point is described by a superconformal field theory (SCFT), whose operators have anomalous dimensions that can be obtained
from the scaling form of the Seiberg-Witten curve.

Although many years have passed since their discovery,
relatively little is known about these inherently strongly-coupled SCFTs.
For example, the field theoretical determination of
the central charges $a$ and $c$ of the superconformal symmetry
has long been an open problem,
despite  progress in the understanding the
dynamics of $\cN=1$ SCFTs using the technique of $a$-maximization \cite{IW}.
In order for $a$-maximization  to work, one needs to know all the $\UONE$
symmetries in the infrared as well as their 't Hooft anomalies.
In practice, one calculates anomalies perturbatively
in the ultraviolet and uses
the 't Hooft anomaly matching conditions to infer
the anomalies of the strongly-coupled infrared SCFT.
However, most infrared $\cN=2$ SCFTs have accidental $\UONE_R$ symmetries, which
limit the usefulness of this technique.

Only recently were the central charges of the conformal and flavor symmetry algebras computed in the simplest of these SCFTs \cite{AS,AW,AT}.
The first such computations were made by Argyres and Seiberg \cite{AS}, who conjectured that many one-loop finite $\cN=2$ gauge theories at infinite coupling admit a dual description as SCFTs with weakly gauged flavor groups.  Using this duality, they were able to obtain the central charges $k$ of the flavor symmetries of certain SCFTs.

Soon after, an alternative, holographic approach was employed to reproduce the flavor central charges of \cite{AS} and to calculate furthermore the conformal central charges $a$ and $c$ 
of an infinite family of SCFTs \cite{AT}.  This calculation was based on the observation that the low-energy theory on $N$ D3-brane probes of an F-theory background with six 7-branes of appropriate charges is equivalent to the $\cN=2$ $\USp(2N)$ gauge theory with four fundamental flavors and one hypermultiplet in the antisymmetric tensor representation
\cite{S,BDS,DLS,ASYT}.   The well-known rank-one superconformal theories of  \cite{APSW,MN1,MN2} and their higher-rank generalizations can all be realized by probing F-theory singularities of Kodaira type, due to collisions of mutually nonlocal 7-branes.
This is the simplest family of SCFTs that admits a large-$N$ description, and which can therefore be studied using holographic techniques \cite{FS,AFM}.

Up to now, the central charges of $\cN=2$ 4d SCFTs have been computed in somewhat indirect ways, by appealing to S- or AdS/CFT duality. The usefulness of such approaches is limited to those SCFTs related by S-duality or holography to weakly coupled theories.
Our main concern in this paper will be to develop a more direct and general method for computing central charges, which can in principle be applied to any $\cN=2$ SCFT
that corresponds to a singular point in the moduli space of an $\cN=2$ gauge theory.
Given  the definition of the conformal central charges as the response of the SCFT
to an external metric perturbation,
$\vev{T^\mu_\mu}\sim c(\text{Weyl})^2 + a(\text{Euler})$, a direct approach to
computing $a$ and $c$ would be to put the $\cN=2$ theory on a curved background space.
The best method for this purpose, valid for any four-manifold background, is to perform a topological twist of the $\cN=2$ supersymmetry \cite{TQFT}. We will find that
the twisting procedure leads to a technique for obtaining the central charges $a$, $c$ and $k$ of any SCFT realized as a fixed point of an $\cN=2$ gauge theory.  We should emphasize that, although our computations are performed in the context of the topologically twisted theory, we will obtain by this method the central charges of the {\it untwisted} theory.

The plan of the paper is as follows:
we will begin in Sec.~\ref{anomaly}
by recalling the relationship between the central charges in question and the $\UONE_R$ anomaly.  Specifically, the divergence of the $R$-current is equal to a sum of topological densities with coefficients equal to linear combinations of central charges \cite{Anselmi1,Anselmi2,KT}.  Topological twisting \cite{TQFT} modifies these coefficients in a simple way, and reduces our problem to computing the $\UONE_R$ anomaly in the associated topological theory.

In Sec.~\ref{holomorphy}, we will see that the $\UONE_R$ anomaly
of the vacuum can be obtained from the $R$-charge of the so-called measure factor
$A^\chi B^\sigma$ in the $u$-plane integral, which has been studied extensively
in the context of Donaldson-Seiberg-Witten theory \cite{W,MW,LNS1,MM,MMP}.
We will recall how these terms were originally determined
from  monodromies and
scaling behavior of codimension-one singularities in moduli space.
We will extend and adapt these results for use in the following sections.

In Sec.~\ref{examples}
we will study  the superconformal theories
realized at the maximally singular points
in the moduli spaces of  $\cN=4$ $\SU(N)$ gauge theory and  $\cN=2$
$\SU(N)$ gauge theory  with $2N$ massless fundamental quarks.
In these theories the coupling is arbitrary, and the central charges
can alternatively be computed in the free-field limit, giving a consistency check
on our method.
Next, we will evaluate the central charges of  the SCFT corresponding to
the maximally singular point in the moduli space of the pure $\SU(N)$ gauge theory.
Finally,  we will study the conformal and flavor central charges for the SCFTs that arise for
gauge group $\SU(2)$ with quarks in the fundamental representation,
and for their generalizations to $\USp(2N)$.
In the large-$N$ limit, these have known gravity duals,
and our results for the central charges in all cases agree with the values
obtained holographically.

In Sec.~\ref{features} we discuss a few general features
of $\cN=2$ SCFTs which can be deduced from our approach.
We will derive the relation between
the  combination $2a-c$ of the central charges
and the sum of the dimensions of the operators parameterizing the Coulomb branch,
originally conjectured by Argyres-Wittig \cite{AW}. We also derive
upper and lower bounds on the ratio $a/c$, verifying a recent conjecture of Hofman and Maldacena \cite{HM}.

\section{Central Charges and Anomalies}\label{anomaly}

We begin by recalling some definitions and basic facts about the flavor and conformal central charges of 4d superconformal field theories.  We review the dependence of the $\UONE_R$ anomalies on the central charges, and discuss how this dependence is modified by topological twisting of $\cN=2$ theories.

\subsection{Definitions and basic relations}

The central charges $a$ and $c$ of conformal symmetry in four dimensions
are defined in terms of operator product expansions (OPEs)
of energy-momentum tensor operators,
but are more elegantly expressed as coefficients of terms in the conformal anomaly of the trace of the energy-momentum tensor
generated by a background gravitational field,
\begin{equation}
\vev{T_\mu^\mu} = {c\over 16\pi^2}(\text{Weyl})^2 - {a\over 16 \pi^2}(\text{Euler})
\end{equation}
where
\begin{eqnarray}
({\rm Weyl})^2 &=& R^2_{\mu\nu\rho\sigma}-2R^2_{\mu\nu}+{1\over 3} R^2,\\
({\rm Euler})&=& R^2_{\mu\nu\rho\sigma}-4R^2_{\mu\nu}+ R^2.
\end{eqnarray}

The $\cN=1$ superconformal algebra relates $a$ and $c$ to the $\UONE_R$ anomalies \cite{Anselmi1,Anselmi2}
\begin{equation}
a={3\over32}\left[3\tr (R_{\cN=1}^3) -\tr (R_{\cN=1})\right],\qquad
c={1\over32}\left[9\tr (R_{\cN=1}^3) -5\tr (R_{\cN=1})\right]\label{a-c-from-r}
\end{equation}  where the trace is over all species of Weyl fermions,
and $R_{\cN=1}$ is the generator of $\UONE_R$ symmetry in the
$\cN=1$ superconformal algebra.
For example, by summing over component fields we find \begin{equation}
a=\frac{1}{24},\qquad c=\frac{1}{12} \label{freehyperAC}
\end{equation} for a free $\cN=2$ full hypermultiplet,
and we have \begin{equation}
a=\frac{5}{24},\qquad c=\frac{1}{6}   \label{freevectorAC}
\end{equation} for a free $\cN=2$ vector multiplet.

The relations \eqref{a-c-from-r} can alternatively be written as an anomaly equation for the $\UONE_R$ current\begin{equation}
\partial_\mu R^\mu_{\cN=1}=
\frac{c-a}{24\pi^2}
R_{\mu\nu\rho\sigma}\tilde R^{\mu\nu\rho\sigma}
+
\frac{5a-3c}{9\pi^2}V_{\mu\nu}^{\cN=1}\tilde V^{\mu\nu}_{\cN=1}.
\end{equation} Here, $V_{\mu\nu}^{\cN=1}$
is the field strength of the external gauge field coupling to the $\UONE_{R,\,\cN=1}$
current.
The Riemann tensor with one tilde is defined as \begin{equation}
\tilde R_{\mu\nu\rho\sigma}=\half\epsilon_{\mu\nu}{}^{ab} R_{ab\rho\sigma}.
\end{equation}

Our focus will be on $\cN=2$ SCFTs, for which the  $R$-symmetry $\UONE_R \times \SU(2)_R$
is related to the  $\UONE_R$ symmetry of its $\cN=1$ subalgebra by
\begin{equation}
R_{\cN=1}=\frac13 R_{\cN=2}+\frac43  I_3,
\end{equation}
where  $I_a$ ($a=1,2,3$) are generators of $\SU(2)_R$,
such that $I_3$ has  eigenvalues $\pm1/2$ in the doublet representation.

For scalar primary operators $\cO$, the $\cN=2$ $R$-charge is related to the dimension by \begin{equation}
R_{\cN=2}(\cO)=2D(\cO).
\end{equation}
Unitarity requires \cite{Mack,DP} the dimension of a scalar primary operator to satisfy
\begin{equation}
D(\cO)\ge 1.
\end{equation} The inequality is saturated only for free fields.
Other basic properties
of $\cN=2$ SCFTs can be found in \cite{APSW}.

The three-point correlators of the $\cN=2$ supercurrents are
known to contain only two superconformal invariants \cite{KT}.
Therefore, we can find the relation of $a$ and $c$  to
the anomalies of the ${\cal N}=2$ $R$-symmetries by
considering free fields. We find
\begin{equation}
 \tr (R^3_{\cN=2})=\tr (R_{\cN=2})=48(a-c),\quad
 \tr (R_{\cN=2}I_a I_b) =\delta_{ab}(4a-2c).
\end{equation}
These conditions translate into an anomaly equation for the $\cN=2$ $R$-current \begin{equation}
\partial_\mu R^\mu_{\cN=2}=
\frac{c-a}{8\pi^2}
R_{\mu\nu\rho\sigma}\tilde R^{\mu\nu\rho\sigma}
+
\frac{3(a-c)}{\pi^2}V_{\mu\nu}^{\cN=2}\tilde V^{\mu\nu}_{\cN=2}
+
\frac{2a-c}{8\pi^2}W_{\mu\nu}^{a}\tilde W^{\mu\nu}_{a}\label{cN=2-relations}
\end{equation}  in the presence of a background metric and a background $\SU(2)_R$ gauge field
$W_{\mu\nu}^a$.

The central charge of an internal global symmetry group $G$
is defined to be  the coefficient $k_G$ of the leading term in the OPE of two $G$-currents
\begin{equation}
J_\mu^a(x)J_\nu^b(0)={3k_G\over 4\pi^4}\delta^{ab}{x^2 g_{\mu\nu}-2x_\mu x_\nu \over x^8}+\cdots . \label{currentcentralcharge-definition}
\end{equation}
We normalize the generators $T^A$ of $G$ so that they have eigenvalues $\pm1$
in the adjoint representation, and
 $k_G$ is normalized so that a Weyl spinor in the fundamental
representation of $\SU(N)$ contributes $1$ to it, as in \cite{AS}.

If $G$ commutes with the supercharges, we call it a flavor symmetry.
$\cN=2$ supersymmetry then relates the current algebra central charge
$k_G$ to the 't Hooft anomaly via the relation
\begin{equation}
k_G\delta^{AB}=-2\tr (R_{\cN=2} T^A T^B).
\end{equation}
This formula is not applicable to the $R$-currents which are in the
superconformal algebra. Instead, the corresponding central charges are
proportional to $c$ because the $R$-currents are in the same supermultiplet
as the energy momentum tensor. The proportionality constants can be
determined by considering free fields, and we have
\begin{equation}
k_{\SU(2)_R}=2c,\qquad
k_{\UONE_R}=16c.
\end{equation}
A non-Abelian flavor central charge leads to an additional term in the anomaly equation for the $\cN=2$ $R$-current  \eqref{cN=2-relations}
\begin{equation}
\partial_\mu R^\mu_{\cN=2} = \cdots -
\frac{k_G}{32\pi^2} F^A_{\mu\nu}\tilde F_A^{\mu\nu}, \label{flavoranomaly}
\end{equation}
where now $F^A$ is the field strength of an external flavor symmetry gauge field.
Henceforth, we will drop the $\cN=2$ subscript on $R^\mu$.

\subsection{Topological twist and central charges}

$\cN=2$ supersymmetric gauge theories are related via a well-known twisting procedure to topological field theories.  In this section, we will demonstrate that the $R$-symmetry anomalies of  $\cN=2$ SCFTs
are closely related to anomalies of their topologically twisted cousins.   Specifically, the coefficients in the anomaly equation are, as in \eqref{cN=2-relations}, linear combinations of $a$ and $c$.  This observation will allow us to compute $a$ and $c$ using the well-developed technology of topological field theories.

The topological twist of an $\cN=2$ gauge theory is performed by introducing an
external $\SU(2)_R$ gauge potential and setting it equal to the
self-dual part of the spin connection.
Since the supercharges transform as a doublet
both under the $\SU(2)_R$ symmetry and under the self-dual
part of the Lorentz group, this causes one component of the supercharge,
which we call $Q_{BRST}$,  to transform effectively as a scalar.
Interpreting $Q_{BRST}$ as a BRST operator, we
define physical operators to be those operators in the cohomology of $Q_{BRST}$.
It can be shown that their correlators do not depend on the metric of the manifold;
this is the sense in which the theory is topological.
In the following we will assume that the manifold has a spin structure in order to avoid subtleties associated with twisted hypermultiplets.

To implement the twisting procedure, we set the $\SU(2)_R$ field strength equal to the
self-dual part of the curvature \begin{equation}
W^a_{\mu\nu}t^a_{\rho\sigma}=
\half (R_{\mu\nu\rho\sigma}+\tilde R_{\mu\nu\rho\sigma}) \label{twist}
\end{equation} where $t^a_{\rho\sigma}$ are $\SU(2)_R$ generators
in the vector representation of SO$(4)$.
The vectors on which $t^a$ acts are the direct sum of two doublets of $\SU(2)_R$, and
the $t^a$ are normalized to have  eigenvalues $\pm\frac{i}{2}$.
For example,
\begin{equation}
t^3_{\rho\sigma}=\half\hbox{\footnotesize$\begin{pmatrix}
&1&&\\
-1&&&\\
&&&1\\
&&-1&
\end{pmatrix}$}.
\end{equation} Then we have \begin{equation}
W^a_{\mu\nu} \tilde W_a^{\mu\nu} = \half\left[
R_{\mu\nu\rho\sigma}\tilde R_{\mu\nu\rho\sigma}
+R_{\mu\nu\rho\sigma}\tilde {\tilde R}_{\mu\nu\rho\sigma}
\right]
\end{equation} where
the Riemann tensor with two tildes is defined as \begin{equation}
\tilde {\tilde R}_{\mu\nu\rho\sigma}=\quarter\epsilon_{\mu\nu}{}^{ab}
\epsilon_{\rho\sigma}{}^{cd} R_{abcd}.
\end{equation}

Making the replacement \eqref{twist}  in the anomaly equation \eqref{cN=2-relations}, we obtain the anomaly equation for the $\cN=2$ $\UONE_R$ current of the twisted superconformal theory,
\begin{equation}
\partial_\mu R^\mu=
\frac{2a-c}{16\pi^2}R_{\mu\nu\rho\sigma}\tilde {\tilde R}_{\mu\nu\rho\sigma}
+
\frac{c}{16\pi^2} R_{\mu\nu\rho\sigma}\tilde R_{\mu\nu\rho\sigma}.
\end{equation}  Integrating over the four-manifold, we find the anomalous shift in the $R$-charge
(ghost number) of the vacuum\footnote{Our convention is opposite to that of \cite{W}, where
$\Delta R$ denotes the total ghost number of the operators appearing in nonzero correlation functions,
i.e.  $\Delta R_\text{there}=-\Delta R_\text{here}$. }
\begin{equation}
\Delta R =  2(2a-c) \chi+ 3c\,\sigma \label{a-c-anomaly}
\end{equation} in terms of the Euler characteristic $\chi$ and the signature $\sigma$ of the manifold, where \begin{align}
\chi & = \frac{1}{32\pi^2}\int \epsilon_{abcd} R^{ab}\wedge R^{cd}
= \frac{1}{32\pi^2}\int d^4x \sqrt{g} R_{abcd}\tilde {\tilde R}_{abcd},\\
\sigma & = \frac{1}{24\pi^2}\int \phantom{\epsilon_{abcd}}R^{ab}\wedge R^{ab}
= \frac{1}{48\pi^2}\int d^4x \sqrt{g} R_{abcd}\tilde { R}_{abcd}.
\end{align}

In particular, a free full hypermultiplet contributes \begin{equation}
\Delta R=\sigma/4 \label{freehyper}
\end{equation}whereas a free vector multiplet contributes \begin{equation}
\Delta R=(\chi+\sigma)/2.\label{freevector}
\end{equation}

It is known that one can introduce an external gauge field for the flavor symmetry $G$
without breaking topological invariance \cite{LM}.
This leads to an extra term in \eqref{a-c-anomaly}
\begin{equation}
\Delta R =  2(2a-c) \chi+ 3c\,\sigma - k_G\,  n
\end{equation} where $n$ is the instanton number \begin{equation}
n=\frac{1}{32\pi^2}\int d^4x \sqrt{g} F^A_{\mu\nu} \tilde F_A^{\mu\nu}.
\end{equation}

\section{Holomorphy and Central Charges}\label{holomorphy}
We saw in the last section that the central charges of an $\cN=2$ SCFT
are captured
by the dependence of the $R$-anomaly of the vacuum $\Delta R$ on the
topology of the underlying four-manifold, in the topologically twisted theory.
As we will soon see,
$\Delta R$  is encoded in the behavior of the topological partition function
near the superconformal points. These two observations will provide the basis for our method of computing central charges of the untwisted theory, which we will now describe in detail.

\subsection{The measure factor $A^\chi B^\sigma$ and the central charges}
The partition function
of the twisted theory is given by the path integral of the
low-energy Lagrangian, which we know from the Seiberg-Witten solution.
Let $u$ denote collectively a set of gauge- and monodromy-invariant complex
coordinates of the Coulomb branch.
At generic values of $u$, the low energy limit is an Abelian gauge theory,
possibly with neutral hypermultiplets.
The coupling  $\tau_{IJ}(u)$ of the low-energy Abelian theory
enters in the twisted Lagrangian as \begin{equation}
S_\text{twisted}\sim \{Q_{BRST}, V\} + \tau_{IJ}(u) F^I\wedge F^J. \label{twistedLag}
\end{equation} The second term, proportional to the instanton number density, receives contributions from the massive degrees of freedom which have been integrated out.
Since the topological theory is defined on a curved manifold, the massive degrees of
freedom will generate additional similar terms in the low energy Lagrangian, proportional to other topological densities of the external gravity and gauge fields, which we can write schematically as
\begin{equation}
\sim (\log A(u)) \tr R\wedge \tilde R  + (\log B(u)) \tr R\wedge R
+ (\log C(u))\tr F_G \wedge F_G.\label{extraLag}
\end{equation}
$A$, $B$, and $C$ are holomorphic functions of $u$, and
$F_G$ is an external gauge field coupled to a non-Abelian global symmetry $G$.
More precisely we define $A$, $B$ and $C$ to be factors in the path integral measure
through which the path integral depends on the topology of the background as follows: \begin{equation}
Z=\int [du][dq] A^\chi B^\sigma C^n e^{-S_{\text{low energy}}} \label{Z}
\end{equation} where $[du]$ and $[dq]$ stand respectively for the path integral measures for the massless vector multiplets and neutral hypermultiplets that are massless throughout the moduli space.
Topological invariance requires $A$, $B$, and $C$ to be holomorphic,
which makes it possible to determine them from their
behaviors in the large-$u$ limit and near singular loci of moduli space  \cite{W,MW,LNS1},
following the strategy used by Seiberg and Witten to find $\tau_{IJ}(u)$ \cite{SW1,SW2}.

The central idea is to exploit the accidental $R$-symmetry which appears
in regions of moduli space where
there exists a duality frame relative to which the theory is weakly coupled.
In such a frame, the $R$-symmetry is realized in the weak-coupling limit as
the $\UONE_R$ in the superconformal group which acts on the massless free fields, and
the contributions to the $R$-anomaly of the additional states that become massless in this limit are encoded in the scaling behavior of $A(u)^\chi B(u)^\sigma$.

As a simple example \cite{W},
let us consider a locus of complex codimension one, where a single
hypermultiplet charged under a $\UONE$ gauge field becomes massless.
Take $\delta u$ to be a transverse local coordinate,
so that the locus is at $\delta u=0$.
For $\delta u \sim 0$ the dynamics is weakly coupled
in the duality frame relative to which the light hypermultiplet is electrically charged,
and in the $\delta u \to 0$ limit the low energy theory is free and thus
trivially superconformal. In particular,  there is an accidental $R$-symmetry,
and $R(\delta u)=2$.
The contribution of the nearly massless hypermultiplet to the $R$-anomaly of the vacuum, as given in \eqref{freehyper},
is not accounted for by the path integral over $\delta u$.
So the gravitational factor should scale holomorphically as
\begin{equation}
A^\chi B^\sigma \sim (\delta u)^{\sigma/8}\label{monopolebehavior}
\end{equation}
in order to reproduce \eqref{freehyper}.
One can perform similar analyses at other types of codimension-1 loci,
and in the asymptotic region $|u|\to\infty$.

In order to determine the functions $A$, $B$ and $C$ over moduli space, we expect that it will be sufficient to consider their analytic properties near codimension-one singularities (and in the large-$u$ limit), and that their behaviors near higher-order singularities at the intersections of codimension-one singularities will be fully determined by analyticity.

Suppose that $A(u)$ and $B(u)$ have  already been calculated in this manner, and
that along some higher-codimension locus one has a strongly-coupled
superconformal theory in the infrared limit.
We first consider a point in the
Coulomb branch which is close to but not at this superconformal point.
The Coulomb branch vevs introduce a scale to the SCFT and spontaneously break the accidental superconformal symmetry. Below this scale, the low-energy theory is
is just a free system consisting of $r$ vector multiplets and $h$ neutral hypermultiplets, with trivial superconformal symmetry and an $R$-anomaly given by
\eqref{freevector} and \eqref{freehyper}, which is contained in the measures $[du]$ and $[dq]$. Now moving to the strongly-coupled superconformal point, we need to add in the $R$-anomaly due to the extra massless fields, which is equal to the $R$-charge of 
the measure factor $A^\chi B^\sigma$. Therefore the total $R$-anomaly is given as the sum,
\begin{equation}
\Delta R = \chi R(A) + \sigma R(B) + \frac{\chi+\sigma}{2} r + \frac{\sigma}{4} h.
\label{DeltaR}
\end{equation}

By comparing to \eqref{a-c-anomaly}
we obtain  general expressions for $a$ and $c$
\begin{align}
a&=\frac14R(A)+\frac16R(B)+\frac5{24}r+ \frac1{24}h,&
c&=\frac13R(B)+\frac1{6}r+ \frac1{12}h\label{master-formula}
\end{align}
which are valid for any $\cN=2$ SCFT that corresponds to a superconformal point in the Coulomb branch of a 4d $\cN=2$ gauge theory.

The evaluation of $a$ and $c$ at the monopole point
\eqref{monopolebehavior} provides a trivial but instructive example.
There, the low energy theory has $r$ free vector multiplets and $h+1$ free hypermultiplets,
so $a=5r/24+(h+1)/24$ while $c=r/6+(h+1)/12$.  We may then apply
\eqref{master-formula} to obtain  $R(A)=0$, $R(B)=1/4$.
In fact this is essentially how the behavior \eqref{monopolebehavior} was originally determined in \cite{W}.

Now all we need to do is to obtain  $R(A)$ and $R(B)$ at
the superconformal points.
This is easy if we know the explicit forms of $A(u)$ and $B(u)$,
because the superconformal $R$-charges of
the Coulomb branch operators $u_i$ are twice their dimensions,
which in turn can be fixed by the analysis
of the Seiberg-Witten curve and differential.

It could happen that the scaling behavior of the Seiberg-Witten curve would not
be sufficient to determine the dimensions of the $u_i$, if
there were a non-$R$ $\UONE$ symmetry acting on the $u_i$ that could mix with the naively defined $\UONE_R$.  In such situations, the $a$-maximization could be used to determine the correct $\UONE_R$ symmetry and to obtain $R(A)$ and $R(B)$. Fortunately or unfortunately, we will not encounter this interesting
possibility in the present paper. (Note that the extra  $\UONE$ flavor
symmetry at the monopole point does not mix with $U(1)_R$ because it only acts on hypermultiplets.)

\subsection{Determination of the measure factor}
We have reduced the determination of $a$ and $c$ to finding the functions $A(u)$ and $B(u)$.
The analysis of \cite{W,MW,LNS1,MM,MMP}
strongly suggests that
\begin{equation}
A=\alpha\left[\det \frac{\partial u_i}{\partial a^I}\right]^{1/2},\qquad
B=\beta\Delta^{1/8}\label{generalconjecture}
\end{equation}
for generic gauge theories.
Here, $u_i$ are gauge- and monodromy- invariant coordinates
on the Coulomb branch, $a^I$ are special coordinates,
and $\Delta$  is the ``discriminant"  of the Seiberg-Witten curve.
(The reason for the double quotes will be explained shortly.)
$\alpha$ and $\beta$ are prefactors independent of the $u_i$ which
can in principle depend on the mass parameters.

The form of $A$ is partly motivated by the observation that the
path integral measure for the $u$ coordinates
has a modular anomaly
\begin{equation}
[du]_{\text{new}}=[du]_\text{old}
\left[\det\frac{\partial a_{\text{new}}^I}{\partial a_{\text{old}}^I}\right]^{-\chi/2}
\end{equation} under a change in the choice of electric special coordinates \cite{W}, such as would result from transport around a codimension-one singular locus.
To ensure modular invariance of the full path integral
(i.e.~single-valuedness of the integrand under monodromies)
this anomaly must be canceled by the modular transformation of $A^\chi$, and $B^\sigma$ should be invariant.  The conjectured forms of $A$ and $B$ \eqref{generalconjecture}  have precisely the modular weights needed to cancel the modular anomaly.

As for the form of $B$, since all free massless states contribute to the part of the $R$-charge proportional to $\sigma$ (i.e., they all contribute to $c$), $B^\sigma$ must vanish along each locus corresponding to the appearance of a single additional massless state.  $B^\sigma$ should also have no monodromy around such loci, so it should also be proportional to a positive integer power of the vanishing modulus $\delta u$. That is, it must at least be divisible by the mathematical discriminant.

In the following we will study the behavior of $A$ and $B$ as defined in \eqref{generalconjecture},
in the asymptotic region $|u|\to \infty$ and near codimension-1
singular loci. We will see that the form \eqref{generalconjecture}
indeed reproduces the expected $R$-anomaly in these regions.
We will also discuss the determination of $\alpha$ and $\beta$
using  holomorphy, as first outlined in \cite{MMP}.

\subsubsection{Weakly-coupled limit $|u|\to\infty$}\label{asymptotic}

First let us consider
the behavior of $A$ and $B$ near $|u|=\infty$, for the pure $\SU(2)$
gauge theory. This is a weak-coupling limit, in which the
charged  vector multiplets $W$ become infinitely massive.
Due to their contribution to the $\UONE_R$ anomaly, these massive $W$'s do not completely decouple, but leave behind a measure factor $A(u)^\chi B(u)^\sigma$ after they are integrated out.
According to \eqref{freevector}, each $W$ contributes $(\chi+\sigma)/2$ to the $R$-anomaly of the vacuum.
Therefore the two massive $W$'s of $\SU(2)$ together give
a measure factor that goes like
\begin{equation}
A(u)^\chi B(u)^\sigma\sim u^{(\chi+\sigma)/4}
\end{equation}
as $u\to\infty$, since the $R$-charge of $u$ in this limit is 4.

More generally, let us consider an $\cN=2$ gauge theory
with gauge group $G$, with $\tilde h$ massive charged hypermultiplets. Each hypermultiplet will  contribute $\sigma/4$ to the $R$-anomaly, according to \eqref{freehyper}. For simplicity we assume that there are no massless hypermultiplets at generic points of the Coulomb branch. Letting $|G|$ and $r$ denote the dimension and the rank of the gauge group, there are $|G|-r$ massive charged vector multiplets in this limit, each of which contributes $(\chi + \sigma)/2$ to $\Delta R$. It then follows that
\begin{equation}
R(A)= \half (|G|-r), \qquad \label{Alimit}
R(B)= \half(|G|-r)+\quarter\tilde h.
\end{equation}

Now let us check that the form of $A$ given in \eqref{generalconjecture}
has the correct behavior as $|u|\to \infty$.  In this limit,
the natural gauge-invariant coordinates $u_i$ are identified with
the vevs $\vev{\tr \phi^{D_i}}$, where
$D_i$ is the degree of the $i$-th Casimir invariant of the group $G$,
and $\phi$ is the adjoint scalar field in the vector multiplet.
In terms of these the dimension of $A$ is
\begin{equation}
D\left(\left[\det \frac{\partial u_i}{\partial a^I}\right]^{1/2}\right)
=\half\sum ( D_i - 1 ).
\end{equation}
Using the relation
\begin{equation}
|G|= \sum_{i=1}^r (2D_i - 1)\label{group-theory-relation}
\end{equation}
we have \begin{equation}
D(A)=  \quarter (|G| - r)
\end{equation} in the $u\to\infty$ region, in agreement with \eqref{Alimit}.

Using \eqref{Alimit} we can now compute the central charges $a$ and $c$ in the
$|u|\to \infty$  limit.  In this free-field limit there are $r$ massless vector multiplets and no massless hypermultiplets, and so
by \eqref{DeltaR} the total $R$-charge of the vacuum is
\begin{equation}
\Delta R_{\infty}
= \half |G|(\sigma + \chi) + \quarter \tilde h\, \sigma\, .
\end{equation}
This precisely matches the anomaly of the underlying microscopic theory, which contains $|G|$ vector multiplets and $\tilde h$ hypermultiplets.
From \eqref{master-formula} we also have
\begin{equation}
a_{\infty}=\frac5{24}|G|+\frac1{24}\tilde h;\qquad
c_{\infty}=\frac1{6}|G|+\frac1{12}\tilde h. \label{acGh}
\end{equation}

Note in particular that
\begin{equation}
4(2a-c)_{\infty}= |G| = \sum_{i=1}^r (2D_i - 1)
\end{equation} from \eqref{group-theory-relation} and \eqref{acGh},
irrespective of the number of the hypermultiplets.
We shall have more to say about the generalization of this equation to other 4d superconformal theories in Sec.~\ref{2a-c}.

\subsubsection{Weakly-coupled codimension-one loci}
\label{codimension-one}
We will now study in more detail the behavior near the locus $\delta u=0$
where a single charged hypermultiplet
becomes massless. We choose the duality frame where the low energy dynamics is
weakly coupled, and $D(\delta u)=1$.
Then as argued in the previous section, we have
\begin{equation}
A^\chi B^\sigma \sim (\delta u)^{\sigma/8}. \label{ABhyper}
\end{equation}
Therefore $A$ should not have a zero, and $B^8$ should have a
first-order zero at $\delta u = 0$.
The formula \eqref{generalconjecture}
passes this test. Indeed,
$D(\delta u) = D(a)$ in the frame where the dynamics is weakly coupled,
and $A(\delta u)\sim (\partial \delta u/\partial a)^{1/2}$
is regular and nonzero there.

The interpretation of the formula for $B$ is a little more subtle.
In general, $B^8$ will have a zero
whenever a hyper- or vector multiplet becomes massless.
When this happens, the integral of the Seiberg-Witten 1-form
around a 1-cycle vanishes. This often (but not always) corresponds to the
vanishing of a period of the Riemann surface.
The mathematical discriminant of a family of Riemann
surfaces $C(u_i)$, parameterized by $u_i$, is defined so as
to have a
zero wherever $C(u_i)$ is singular; that is, wherever one of its cycles degenerates.
Thus $B^8$ behaves in a very similar way
to the mathematical discriminant.
However, there are differences: first, it is possible to associate to a gauge theory
more than one family of Seiberg-Witten curves, with different mathematical discriminants.  This is the case in the original example of pure $\SU(2)$  \cite{SW1,SW2}, where
two possible curves for $\SU(2)$ share the same zero locus,
but the orders of the zeroes are different.
Second, for a general gauge theory, not
all of the cycles of the Seiberg-Witten curve necessarily correspond to physical states.  Therefore some of the zeroes of the mathematical
discriminant may not correspond to the appearance of extra massless degrees of freedom.
Still, the form of $B^8$ we will find is very close to the mathematical
discriminant. Therefore,  we will define the
{\it physical discriminant} $\Delta \equiv B^8$ to be an object which has a zero wherever an additional state becomes massless. If that state is a single hypermultiplet, the zero should be first-order, to reproduce \eqref{ABhyper}.

Another type of singularity that we will encounter corresponds to an
enhanced $\SU(2)$ gauge symmetry accompanied by enhanced $\cN=4$ supersymmetry.
At such a singularity two $\cN=4$ vector multiplets become massless, in addition to an  $\cN=4$ $\UONE$ vector multiplet that is massless everywhere.
There are two
natural coordinates near the singularity, $\delta u$ and $a$, with
$\delta u\sim a^2$ as the singularity is approached.
$a$ has dimension one but is not monodromy invariant;
$\delta u$ is invariant but has dimension two.
In terms of $a$ we have \begin{equation}
A^\chi B^\sigma \sim a^{\chi/2+3\sigma/4}. \label{ABN=4}
\end{equation}
Note that $A$  scales as  $\sim a^{1/2}$, in agreement with the
identification $A=(\partial (\delta u)/\partial a)^{1/2}$ of \eqref{generalconjecture}.
Also, according to \eqref{ABN=4} the physical discriminant
$\Delta=B^8$ should scale as $a^6$.
To compare that with the mathematical discriminant of the curve of the
$\cN=4$ $\SU(2)$ theory, recall that the curve 
is \begin{equation}
y^2=x^3-g_2(\tau) xu^2 -g_3(\tau) u^3
\end{equation} where $g_{2,3}(\tau)$  are the usual Eisenstein series up to
a constant factor. The mathematical discriminant scales as $u^6\sim a^{12}$, so
in this case, the physical discriminant is the square root of the mathematical one.

More generally, suppose in a given weakly-coupled limit an
extra $\tilde r$ vector multiplets and $\tilde h$ neutral hypermultiplets become
massless. (These are in fact
the only supermultiplets that can be massless in an $\cN=2$ theory.)
The theory is trivially superconformal, with an accidental $R$-symmetry which
acts on the extra massless states as well as on the $r$  free vectors and the $h$ free hypermultiplets that are massless throughout the moduli space.
The $R$-anomaly of $r$ free vector multiplets and $h$ free hypermultiplets
is accounted for by the path integral measures $[du]$ and $[dq]$,
and the factors $A(u)$ and $B(u)$ should reproduce
the anomaly from the accidental $R$-symmetry of the nearly massless fields.
Applying \eqref{freehyper} and \eqref{freevector} we find
\begin{equation}
R(A)=\tilde r /2,\qquad
R(B)=\tilde r /2+\tilde h /4. \label{the-result}
\end{equation}
We expect that the formula for $A$ \eqref{generalconjecture}
will reproduce this $R(A)$, while we define the physical discriminant
$\Delta=B^8$ in such a way that the equation for $R(B)$ is also satisfied. Specifically, $\Delta$ should have a zero of order $(2\tilde r+\tilde h)$.

\subsubsection{The prefactors $\alpha$ and $\beta$}
Before moving on to applications of these ideas to specific SCFTs,
let us briefly consider how the prefactors $\alpha$ and $\beta$ in \eqref{generalconjecture}
can be determined.  The properties of $\alpha$ and $\beta$ were 
first studied in \cite{MMP}, where it was noted that
$\alpha$ and $\beta$ may be functions of the masses $m_a$ of the hypermultiplets,
but are independent of the $u_i$.
If $\alpha$ or $\beta$ develops a pole or a zero at some particular set of
masses $m_a^*$, then new physics must be present at $m_a=m_a^*$ for every
value of $u_i$, i.e. everywhere on the Coulomb branch.

Consider  for example the $\SU(2)$ gauge theory  with two fundamental hypermultiplets
with different masses $m_1$ and $m_2$.
A Higgs branch appears when $m=m_1=m_2$; generically, it touches the Coulomb
branch at a single point $u=u(m)$.
The existence of a Higgs branch should not strongly affect
the physics far away from $u=u(m)$, so in this case we do not expect
that $\alpha$ and $\beta$ will have any mass dependence.
As another example, consider the $\SU(2)$ gauge theory with
one adjoint hypermultiplet with mass $m$. In this case, one component
of the adjoint hypermultiplet has a mass $m$ independent of $u$.
Therefore, in the limit $m\to 0$ there is a single nearly-massless
hypermultiplet irrespective of $u$,
whose anomaly needs to be accounted for by the $\alpha^\chi \beta^\sigma$ factor.
From \eqref{the-result}, $R(\alpha)=0$ and $R(\beta)=1/4$, while
$R(m)=2$. Thus we conclude \begin{equation}
\alpha^\chi \beta^\sigma \sim m^{\sigma/8},
\end{equation} which reproduces the result found in \cite{LL} from a
different perspective.

In the examples below, one may easily check that
$\alpha$, $\beta$, and the corresponding
factor for the external gauge field are all independent of the mass parameters.
This can be done by means of the usual combination of $R$-symmetry and holomorphy arguments, so we will not discuss them further.

\section{Examples}\label{examples}

We now apply our general discussion in the previous sections to calculate the central charges of various SCFTs. In cases for which gravity duals are known, we will find our results consistent with results previously obtained using the AdS/CFT duality.

\subsection{Finite $\cN=2$ theories}
As a first example let us check that our method
correctly reproduces  the central charges $a$ and $c$
of some of the $\cN=2$ theories with vanishing beta functions,
i.e. those theories with an exactly marginal coupling $\tau$.
In such theories, the central charges cannot depend on $\tau$ \cite{AS},
so they can be determined
in the weak coupling limit by simply counting the number of multiplets.
We will show that our formalism gives identical answers, and thus obtain a nice consistency check.

\subsubsection{$\cN=4$ $\SU(N)$ theory}
The $\cN=4$ $\SU(N)$ super Yang-Mills theory can be thought of as an $\cN=2$ $\SU(N)$
gauge theory with one hypermultiplet in the adjoint representation.
The Coulomb branch is parameterized by the vevs of the scalar components  of the
unbroken $\UONE^{N-1}$ subgroup,
$a_i$ for $i=1,2,\ldots,N$.  The special coordinates $a_i$ are subject to the traceless constraint $\sum_i a_i=0$ and are to be identified under the Weyl exchange $a_i\leftrightarrow a_j$. A gauge-invariant alternative set of coordinates is provided by the Casimir operators $u^{(k)}=\vev{\tr \phi^k}= \sum_i a_i^k$, for $k=2,3,\ldots,N$, constructed from the adjoint scalar field $\phi$
in the vector multiplet.
In the superconformal limit, $a_i \to 0$ with mass dimension one; thus $u^{(k)}$ has  $R$-charge 2$k$, and we have \begin{equation}
R(A)=\sum_{k=2}^N (k-1)=\half N(N-1).\label{RA}
\end{equation}

To evaluate $B$, we note that the codimension-one singularities of moduli space
are due to the $\cN=4$ $\SU(2)$ enhancements at $a_i=a_j$ for every pair $i,j$.  Counting the $R$-charges of the two additional massless hypermultiplet and vector multiplet states, we find that \begin{equation}
B^8=\Delta=\prod_{i>j} (a_i-a_j)^6.
\end{equation} Therefore we have \begin{equation}
R(B)=\frac34 N(N-1).
\end{equation} The numbers $r,h$ of free vector- and hypermultiplets
are both $N-1$, so from the formula \eqref{master-formula}
we conclude that\begin{equation}
a=c=\quarter(N^2-1)=\quarter |G|\ ,
\end{equation} which is exactly what we get from
the free field values \eqref{freehyperAC}, \eqref{freevectorAC}.
It is easy to generalize this argument to any $\cN=4$ theory by using \eqref{group-theory-relation}.

\subsubsection{$\cN=2$ $\SU(N)$ theory with $N_f=2N$ quarks}
Let us next study the slightly less trivial example of
$\cN=2$ $\SU(N)$ gauge theory with $2N$  hypermultiplets
in the fundamental representation.
The Seiberg-Witten curve is given by \cite{APS} \begin{equation}
y^2 = P(x)^2- f(\tau) Q(x),
\end{equation}where \begin{align}
P(x)& = x^N+u_{(2)}x^{N-2} + u_{(3)} x^{N-3}\cdots +u_{(N)},
\label{Px}\\
Q(x)&=\prod_{a=1}^{2N} (x-m_a - 2g(\tau)\mu).
\end{align} Here,
$\mu = (1/N_f)\sum m_a$, and $f(\tau)$ and $g(\tau)$ are certain modular functions of the complexified gauge coupling $\tau$.
We take generic masses $m_a$, which will be sent to zero later.
The curve is a hyperelliptic curve of genus $N-1$, obtained by
attaching two copies of the $x$-plane along $N$ cuts.

As in the last example, $R(A)$ is given by \eqref{RA}.
As for $R(B)$,
codimension-one singularities in the moduli space occur when two
solutions of $F(x)\equiv P(x)^2-f(\tau)Q(x)=0$
collide, making the hyperelliptic curve degenerate. Physically, these singularities are all due to a hypermultiplet, electrically charged under a $\UONE$ factor of the low-energy
gauge group, becoming massless. 
Denoting the $2N$ zeros of $F(x)$ by $e_\alpha$, $\alpha=1,2,\ldots,2N$,
we thus have \begin{equation}
B^8=\Delta=\prod_{\alpha>\beta}(e_\alpha-e_\beta)^2.
\end{equation} Superconformal symmetry is restored
in the limit $m_a\to 0$, $u_{(k)}\to 0$. There, the scaling dimensions of
the operators are equal to their mass dimensions. In particular, the dimension of
$e_\alpha$ is one. Thus we have \begin{equation}
R(B)=\quarter 2N(2N-1).
\end{equation}

At a generic point on the Coulomb branch, there are $r=N-1$
free massless vector multiplets and no massless hypermultiplets. Hence we obtain
\begin{align}
a&
= \frac7{24}N^2-\frac5{24},&
c&
=\frac13 N^2-\frac16. \label{error}
\end{align} from our formula \eqref{master-formula}.

\subsection{$\SU(N)$ Argyres-Douglas points}\label{ADac}
After these warm-up exercises,
we will now move on to inherently strongly-coupled superconformal theories.
We will first consider a well-known infinite series of superconformal
points, which generalizes the original $\SU(3)$ Argyres-Douglas point \cite{AD}
to $\SU(N)$ \cite{EHIY}.
Specifically, consider  the $\cN=2$ supersymmetric $\SU(N)$ gauge theory without matter hypermultiplets, with Seiberg-Witten curve \begin{equation}
y^2=P(x)^2 - \Lambda^{2N}
\end{equation}  where $\Lambda$ is the dynamical scale of the theory,
and $P(x)$ is again the degree-$N$ polynomial in $x$
with coefficients $u_{(k)}$ as before.
The Seiberg-Witten differential is \begin{equation}
\lambda=  2 \frac{xdP}y.  
\end{equation}

The Argyres-Douglas point in question is reached by taking \begin{equation}
u_{(2)}=u_{(3)}=\cdots=u_{(N-1)}=0, \qquad u_{(N)}=\Lambda^{2N}
\end{equation} so that the curve becomes \begin{equation}
y^2=x^N(x^N+2\Lambda^{N}). \label{ADcurve}
\end{equation}
The deformations away from this point are parameterized by
\begin{equation}
\tilde u_i=u_{(i)} \ (i=2,\ldots,N-1),\qquad\tilde u_N=u_{(N)}-\Lambda^N.
\end{equation} The Seiberg-Witten differential behaves as \begin{equation}
\lambda 
\sim \frac {2x}{y} d\left[x^N+\tilde u_2 x^{N-2}+\cdots+\tilde u_N\right].
\end{equation} Demanding scaling and $D(\lambda)=1$,
we find \begin{equation}
D(x)=\frac{2}{N+2},\quad
D(y)=\frac{N}{N+2},\quad
D(\tilde u_i)=\frac{2i}{N+2}.
\end{equation}

Note that \begin{equation}
D(\tilde u_{i+1})+D(\tilde u_{N-i+1})=2, \qquad (i=1,2,\ldots,r)
\end{equation} where $r=\lfloor (N-1)/2\rfloor$.
We introduce the notations \begin{align}
\cO_j&=\tilde u_{N-j+1} &&\text{for $j=1,\ldots,r$},\\
\mu_j&=\tilde u_{j+1} &&\text{for $j=1,\ldots,N-1-r$}.
\end{align}  The labels are chosen so that $D(\cO_j)>1$ and $D(\mu_j)\le 1$.
As discussed in \cite{APSW},
$\cO_i$
are local operators in the conformal theory obeying the
unitarity bound,
and $\mu_i$ are the corresponding deformation parameters.
The product of $\mu_i$ and $\cO_i$ is a dimension-two operator which can
be added to the Lagrangian as a deformation   \begin{equation}
\sim \int d^4\theta\, \mu_i\cO_i.
\end{equation} We also have $D(\tilde u_r)=1$
for $N=2r+2$. From the unitarity bound,
the dimension of a scalar operator is greater than one unless it is free,
in which case it has
dimension one. Thus the scalar field
$\tilde u_r$ is free; and the supermultiplet containing it is also free and decouples
from the rest of the theory.  We conclude that the dimension
of the Coulomb branch of the maximal-rank Argyres-Douglas
points for the pure $\cN=2$ $\SU(N)$ gauge theory  is given by
\begin{equation}
r=\lfloor \frac{N-1}{2}\rfloor.
\end{equation}

\begin{figure}
\[
\begin{array}{ccc}
\includegraphics[width=.3\textwidth]{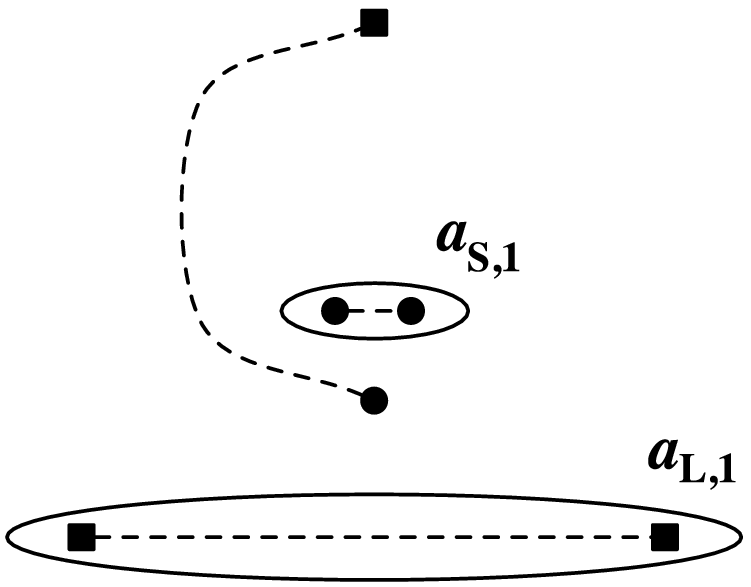} &\qquad\qquad &
\includegraphics[width=.33\textwidth]{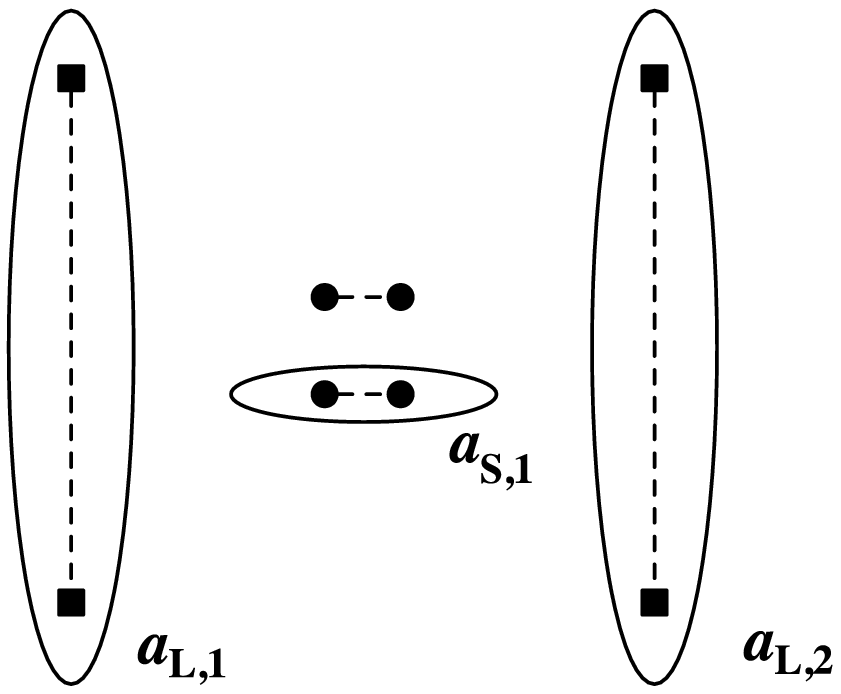}  \\
\SU(3) && \SU(4)
\end{array}
\]
\caption{Cuts and cycles for $\SU(N)$ Argyres-Douglas points.\label{adcuts}}
\end{figure}

The dimension of the Coulomb branch can also be checked from
the structure of the cuts on the $x$-plane, as
depicted in Fig.~\ref{adcuts}.
There, black disks denote branch points near $x\sim 0$,
and black squares denote  those around $x\sim O(\Lambda)$.
We have $r$ short cuts near $x=0$, $r$ long cuts with $x\sim O(\Lambda)$,
and one extra cut connecting a branch point with $x\sim 0$ to a branch point $x\sim O(\Lambda)$ if $N$ is odd.
Therefore we see that $r$ of the special coordinates $a_{S,1},\ldots,a_{S,r}$ are $\sim 0$,
while the rest $a_{L,1},\ldots,a_{L,N-1-r}$
are $O(\Lambda)$.  Correspondingly, at this point $r$ electrically charged states are becoming massless.

To calculate $R(A)$, we first determine the dimension of
$\det (\partial a_I/\partial \tilde u_i)$. It is important to note that \begin{equation}
D(a_{S,i})=1, \qquad (i=1,2,\ldots, r)
\end{equation}while the large special coordinates behave as \begin{equation}
a_{L,i} = \text{constant} + \text{analytic functions of $\tilde u_i$}
\qquad (i=1,2,\ldots,N-1-r),
\end{equation} because the Seiberg-Witten differential
near the long cuts can be expanded in powers of $\tilde u_i$.
The expansion of the
Jacobian $\det (\partial a_I/\partial \tilde u_i)$
has many terms,
and it is easy to see that
the term with the largest scaling dimension is given by
\begin{equation}
\left(\det \frac{\partial a_{S,i}}{\partial\cO_j} \right)
\left( \det \frac{\partial a_{L,i}}{\partial\mu_j}\right)\label{ADleading}
\end{equation}
where the $\cO_i$
differentiate the small special coordinates $a_{S,i}$
and  the $\mu_i$ differentiate the large special coordinates
$a_{L,i}$, if this term does not accidentally vanish. We prove
that the first factor in \eqref{ADleading} is  nonzero, and
that the second factor approaches a nonzero constant  in Appendix~\ref{ADproof}.
Therefore
\begin{equation}
R(A)=2D(A)=\sum_{i=1}^r (D(\cO_i)-1)=
\left\{
\begin{array}{cl}
\displaystyle\frac{r^2}{2r+3}&\quad {(N=2r+1),}\\[3mm]
\displaystyle\frac{r(r+1)}{2r+4}&\quad{(N=2r+2).}
\end{array}
\right.\label{2a-c:SU(n)-AD}
\end{equation}

The discriminant is given by \begin{equation}
B^8=\Delta\sim \prod_{i<j} (x_i-x_j)^2
\end{equation} where $x_i$ $(i=1,\ldots,N)$ are the solutions of \begin{equation}
x^N+\tilde u_2 x^{N-2}+\cdots+\tilde u_N=0.
\end{equation}
Hence we have \begin{equation}
R(B)=\frac14D(\Delta)=\frac14 N(N-1)D(x)=\frac{N(N-1)}{2(N+2)}.
\end{equation}
As the final data we need the number of the operators parameterizing the Coulomb branch.
The Coulomb branch of the original $\SU(N)$ theory has dimension $N-1$,
but as has been discussed only $r$ of them belong to the interacting conformal theory.
In other words, the infrared limit of the original $\SU(N)$ theory at the point
$\tilde u_2=\cdots=\tilde u_N=0$ is the Argyres-Douglas superconformal
field theory plus $N-1-r$ free $\UONE$ vector multiplets. There are no massless
hypermultiplets at generic points of the moduli space.
Therefore, the central charges of the Argyres-Douglas points
are, from \eqref{master-formula},
\begin{equation}
a=\frac{r(24r+19)}{24(2r+3)},\qquad
c=\frac{r(6r+5)}{6(2r+3)}
\end{equation} for $N=2r+1$ and
\begin{equation}
a=\frac{12r^2+19r+2}{24(r+2)},\qquad
c=\frac{3r^2+5r+1}{6(r+2)}
\end{equation} for $N=2r+2$.
Note that \begin{equation}
a=\frac{43}{120},\qquad  c=\frac{11}{30}\label{su3result}
\end{equation} for $N=3$.

\subsection{Models with F-theory realizations}
As another example we study superconformal theories
with F-theory realizations. Let us first quickly recall the construction
and the matter content of the gauge theory.

We start by placing an O7-plane and $N_f$  parallel D7-branes
in a flat 10d spacetime. We probe this system by one or more
D3-branes, and we would like to understand the low energy dynamics
of the theory on the D3-brane worldvolume.  If probed by a
single D3-brane, the gauge group is $\SU(2)$ and there are $N_f$ hypermultiplets
in the doublet of the gauge group. The geometry transverse
to the D7-branes can be identified with the $u$-plane \cite{BDS}, where as usual $u$
is a gauge and modular invariant complex coordinate on the Coulomb branch of the gauge theory, identified with $\vev{\tr \phi^2}$ in the $|u|\to \infty$ limit,
and the relative positions of the D7-branes determine the mass parameters
of the hypermultiplets.

When we probe the 7-brane system by $N$ D3-branes,
the quantization of open strings gives us the gauge group $\USp(2N)$
with $N_f$ massive
hypermultiplets in the fundamental $2N$-dimensional
representation and one massless
hypermultiplet in the antisymmetric tensor\footnote{
The trace part of the antisymmetric tensor is neutral under the gauge group
so we eliminate it in the following analysis.}.
The $N$ D3-branes can move independently, and each probes the same transverse geometry of the 7-branes, which means that
the Coulomb branch can be parameterized by their locations in the $u$-plane,  $u_i$ $(i=1,\ldots,N)$. The D3-branes are indistinguishable, so the coordinates $u_i$ are identified under interchanges $u_i\leftrightarrow u_j$ for each pair $i<j$ \cite{DLS}.

For $N_f=4$, the theory becomes perturbatively superconformal
when one puts all four D7-branes on top of the O7-plane.
For $N_f<4$, the O7-plane splits into two 7-branes nonperturbatively,
which correspond to the monopole and
the dyon points of the $u$-plane.
A strongly-coupled superconformal point
is reached by placing $N_f$ 7-branes on top of the monopole point.
For $N=1$ and $N_f=1,2,3$, points of this type correspond to the superconformal theories that were studied in \cite{APSW}.
The dimension $D(u)$ of the Coulomb branch operator
satisfies the relation \begin{equation}
\frac{1+N_f}{12}=1-\frac1{D(u)}, \label{deficit}
\end{equation} which in the F-theory realization reflects the deficit angle
created by the stack of 7-branes.
Probing with multiple D3-branes
gives rank-$N$ versions of these SCFTs.

In the large-$N$ limit, one can take the near horizon limit of
$N$ D3-branes and study the properties of these superconformal theories
from the gravity point of view \cite{FS,AFM}. The central charges
$a$ and $c$ for these theories were calculated from this perspective in \cite{AT}.
Here we will recalculate them, both as a test of our theoretical methods and as a check on the prior results.
For simplicity we exclude the case $N_f=4$ in the following.

$R(A)$ and $R(B)$ can be extracted from
the known Seiberg-Witten curve of this theory \cite{AP},
but it is easier to analyze directly the singularities in moduli space.
First let us recall the
Coulomb branch of the theory more fully, for the case $N=1$.
For generic hypermultiplet masses
$m_a$ ($a=1,\ldots,N_f$), there are $2+N_f$ singularities on the $u$-plane
at $u=u_\alpha(m_1,\ldots,m_{N_f})$, $(\alpha=1,2,\ldots,2+N_f)$,
which are given by zeroes of
the discriminant \begin{equation}
\Delta\equiv\Delta_{1}(u;m_1,\ldots,m_{N_f}).
\end{equation} As such, the functions $u_\alpha(m_1,\ldots,m_{N_f})$ have monodromies
exchanging them, and there are no absolute distinctions among them.
Still, if $m_a\gg \Lambda$, there are two zeros with $u\sim  O(\Lambda)$ and
$N_f$ zeros at $u\sim m_a^2$. The former are the points
where a monopole or a dyon becomes massless, and the latter
are where the quarks become massless.
For $N>1$ the Coulomb branch is parameterized by $u_1,\ldots,u_N$,
identified under the exchanges $u_i\leftrightarrow u_j$.
Monodromy invariant coordinates are given by
the $k$-th symmetric polynomials $u^{(k)}$
of the  $u_i$, in correspondence with the Casimirs $\vev{\tr\phi^{2k}}$ of the
$\USp(2N)$ gauge group.

The factor $A$ is then given by \begin{equation}
A=\left[\det{\frac{\partial u^{(k)}}{\partial a_i}}\right]^{1/2}. \label{uspA}
\end{equation}
The discriminant is
\begin{align}
\Delta&= \prod_{i>j} (u_i-u_j)^6 \prod_{i,\alpha} (u_i-u_\alpha(m_1,\ldots,m_{N_f}))\\
&\equiv\prod_{i>j} (u_i-u_j)^6 \prod_{i}\Delta_{1}(u_i;m_1,\ldots,m_{N_f}).\label{uspB}
\end{align}
where the first factor accounts for the enhancement of a single $\UONE$ vector multiplet to an $\cN=4$ $\SU(2)$ multiplet  when $u_i=u_j$, i.e., when two D3-branes collide, and the second factor accounts for the appearance of one massless hypermultiplet when $u_i=u_a(m_1,\ldots,m_{N_f})$.
It is also easy to see that $\Delta$ is, as required, a polynomial in the gauge invariant coordinates $u^{(k)}$ and the masses $m_i$.

A superconformal point is reached if we tune $m_1,\ldots,m_{N_f}$ so that the
$N_f$ ``quark'' zeros of the discriminant collide with the monopole zero.
We shift $u$ by a constant so that the
multiple zero is at $u=0$. Then the discriminant becomes \begin{equation}
\Delta= \prod_{i>j} (u_i-u_j)^6 \prod_i u_i {} ^{1+N_f}.
\end{equation}
Therefore \begin{equation}
R(\Delta)=2D(u) \left[
	(1+N_f) N + 3N(N-1)
\right].
\end{equation} $R(A)$ is also easy to determine, because \begin{equation}
D(u^{(k)})=k D(u)
\end{equation} and all of the $a_i$ behave as dimension-1 operators. Thus we have \begin{equation}
R(A)=\sum_k (k D(u) -1) = \frac12 N(N+1) D(u) -N.\label{2a-c:USp(2n)-AD}
\end{equation}

Finally we need the number $r$ of free vector multiplets and
the number $h$ of free hypermultiplets
at generic points of the moduli space, which are easily found to be
$r=N$ and $h=N-1$. Combining the data, we have \begin{align}
a&= \frac14 DN^2 +\frac1{24}(1+N_f ) DN-\frac1{24},\\
c&= \frac14 DN^2 + \frac1{12} \left[(N_f-2)D+3\right]N  -\frac1{12}
\end{align}where we have abbreviated $D(u)$ as $D$.
 Using the relation \eqref{deficit},
these equations become
\begin{align}
a&=\frac14 DN^2 + \frac12 (D-1)N-\frac{1}{24}, &
c&=\frac14 DN^2 + \frac34 (D-1)N-\frac{1}{12}.\label{uspresult}
\end{align}

For $N=1$ and $N_f=1$,  $D(u)=6/5$ and the central charges
are \begin{equation}
a=\frac{43}{120},\qquad  c=\frac{11}{30},
\end{equation} in agreement with the result \eqref{su3result}.
This is as it should be, because the superconformal points
of the pure $\SU(3)$ theory and of the  $\SU(2)$ theory with $N_f=1$ doublets
are believed to be equivalent \cite{DW}.

The result \eqref{uspresult} also completely reproduces the
central charges calculated in \cite{AT} from the gravity dual.
We find the agreement quite nontrivial. In the holographic approach,
the $\cO(N^2)$, $\cO(N)$ and $\cO(1)$ terms in \eqref{uspresult}
arose as contributions due from classical bulk gravity,  branes,
and one-loop effects, respectively, whereas
in our present approach $a$ and $c$ were calculated nonperturbatively and
received contributions from completely different sources, $R(A)$ and $R(B)$.
Furthermore, our formula \eqref{uspresult} also reproduces the central charges
of rank-$N$ versions of the $E_n$ theories if we use the corresponding
dimensions $D(u)$.
These mysterious theories have yet to be realized in a purely field-theoretical language,
but our result strongly suggests
that their gravitational measure factors $A$ and $B$ should still be given by the general formulas
\eqref{uspA}, \eqref{uspB}.

\subsection{Flavor symmetry}
As a final exercise in this section,
let us calculate the central charge of the flavor symmetry
current algebra of the $\USp(2N)$ theory we considered
in the last section.

The flavor symmetry acting on the hypermultiplets
in the fundamental is $U(N_f)=\UONE\times \SU(N_f)$ when
we take all of the masses to be equal, $m=m_1=\cdots=m_{N_f}$,
while the hypermultiplet in the antisymmetric  is acted on by
the $\SU(2)_L$ symmetry.
We study the response of the gauge theory to the introduction
of an external gauge field for the flavor symmetry,
for a generic value of $m$. Later we will take the superconformal limit
and find the current algebra central charge.
\footnote{
The authors learned after the completion of the paper that
the flavor symmetry central charge
of the rank-1 SCFT with $E_8$ flavor symmetry
was determined in \cite{Ganor}, using basically the same method with ours .
}\ 
We denote the resulting measure factor in the low-energy path integral
by $C^nC_L^{n_L}$, where $n$ and $n_L$ are the instanton numbers
of the external $\SU(N_f)$ and $\SU(2)_L$ gauge fields, respectively.

\subsubsection{$\SU(N_f)$}

Although $\UONE\times \SU(N_f)$ naturally acts on the hypermultiplets
in the fundamental representation of $U(N_f)$,
the analysis of the $\UONE_F$ part is quite subtle because of its mixing
with the physical $\UONE$ gauge fields on the Coulomb branch.
So let us study  the current algebra
central charges of $\SU(N_f)$ for $N_f=2,3$ first.
We will come back to $\UONE_F$ in Sec.~\ref{u1f}.

We begin with the case $N=1$.
As we have discussed, if the masses are generic and unequal,
there are $2+N_f$ singular points in the $u$-plane which are given by the
zeroes of the discriminant $\Delta_{1}$.  
When we take $m=m_1=\cdots= m_{N_f}$, the discriminant has
a zero of order $N_f$, \begin{equation}
\Delta_{1}(u;m,\ldots,m)=\underline\Delta(u,m) (u-u_q(m))^{N_f}.
\end{equation} Here $\underline\Delta(u,m)$ is a quadratic polynomial
whose zeroes give the points where a monopole or a dyon becomes massless.
$u=u_q(m)$ is the point where a hypermultiplet in the
fundamental representation of $\SU(N_f)$ appears.  It is important to note that
$u_q(m)$ is a polynomial in $m$ and has no monodromy.
The physical reason is that there is an $(N_f-1)$-dimensional Higgs branch
emanating from $u=u_q(m)$, so that the singularity there
can be clearly differentiated from the monopole and dyon points.

Let now turn to the rank-$N$ version of the theory, parameterized
by $u_1,\ldots,u_N$ with the identification $u_i\leftrightarrow u_j$.
When $u_i=u_q(m)$, one free massless hypermultiplet
in the fundamental of $U(N_f)$ appears, and it contributes to the $R$-anomaly an amount
\begin{equation}
\Delta R=\cdots - 2n,
\end{equation} so $C\sim (u_i-u_q(m))^{-1}$. Thus  \begin{equation}
C=\prod_i (u_i-u_q(m))^{-1}
\end{equation}  times an extra factor which is a holomorphic function with
neither poles nor zeros.  The factor should be constant,
because the function $C$ as written above
already reproduces the correct anomaly when all of the $u_i$ are large.

$C$ is a well-defined function because $u_q(m)$ is a polynomial for $N_f=2,3$. However, we can also see that if we naively try to generalize our result to $\UONE_F$ for
$N_f=1$ we run into a trouble. In this case, $u_q(m)$ is one of the roots
of a cubic polynomial $\Delta(u,m)$ so that it has branch cuts and monodromies.
Therefore  $C$ is not well-defined as a function.
We will come back to this point in Sec.~\ref{u1f}.

Finally, let us calculate the current algebra central charge.
We choose $m$ so that $u_q(m)$ collides with another zero
of the discriminant, which we take to be at $u=0$.
We then have \begin{equation}
C=\prod_i u_i^{-1}
\end{equation} which means that \begin{equation}
\Delta R=\cdots -2ND(u)n,
\end{equation}i.e. \begin{equation}
k_G=2ND(u).
\end{equation} This result agrees with the holographic calculation of \cite{AT}.

\subsubsection{$\SU(2)_L$}
We now calculate the central charge of the $\SU(2)_L$ flavor
symmetry. The antisymmetric traceless representation of the $\USp(2N)$ gauge group
has $N(2N-1)-1$ components. Each component transforms
as an $\SU(2)_L$ doublet of half-hypermultiplets.

At generic points on the moduli space,
there are $N-1$ free hypermultiplets, which are the components
of the antisymmetric of $\USp(2N)$
left uneaten by the breaking of the gauge symmetry to $\UONE^N$.
Each of these is again an $\SU(2)_L$ doublet of half-hypermultiplets.
When $u_i=u_j$ and $\cN=4$ $\SU(2)$ gauge symmetry is enhanced,
two extra components of the field in the antisymmetric becomes
massless, which provide two more $\SU(2)_L$ doublets of half-hypermultiplets.
Finally, when $u_i=u_\alpha(m_1,\ldots,m_{N_f})$,
one hypermultiplet becomes massless, but it is charged only under
the $\UONE$  gauge symmetry, so the accidental flavor symmetry there
is at most $SO(2)$ and one cannot implement $\SU(2)_L$ symmetry.

Thus we have \begin{equation}
C_L =\prod_{i>j} (u_i-u_j)^{-1}
\end{equation} times an extra factor which is holomorphic without
zeros nor poles. Again this extra factor is constant because $C_L$ has dimension
$N(N-1)$ in the asymptotic region,
which is exactly as needed to account for the anomaly
from the massive components of the field in the antisymmetric.
Recalling the fact that $N-1$ free hypermultiplets also contribute,
the $R$-anomaly of the vacuum is given by \begin{equation}
\Delta R=\cdots- \left[D(u)N(N-1)+(N-1)\right]n_L.
\end{equation}
Therefore we conclude \begin{equation}
k_L=D(u)N(N-1)+(N-1)
\end{equation} which also agrees with what was obtained in \cite{AT}\footnote{
The value stated in v2 of \cite{AT} was off by a factor of two, which came from
a mistake in the normalization. It has been corrected in v3.}.

\subsubsection{$\UONE_F$}\label{u1f}
Let us come back to the subtlety surrounding the $\UONE_F$ symmetry.
The naive generalization of the $\SU(N_f)$ analysis fails, as we have seen,
and the failure occurs even before taking the superconformal limit.
Here we will explain the physical reason for this failure, in general terms
that will apply to any $\cN=2$ theory with a $\UONE$ flavor symmetry.

The low-energy form of the topologically-twisted Lagrangian
\eqref{twistedLag}, \eqref{extraLag} does not include all the possible terms
for $\UONE$ flavor symmetries. Indeed, if we have  physical $\UONE$ gauge fields
$F^I$ and  external $\UONE$ gauge fields $F^a$ coupled to the $\UONE$ flavor
symmetries, we can consider the following structure \begin{equation}
S\sim \{Q_{BRST},V\}+\tau_{IJ} (u) F^I\wedge F^J
+ \kappa_{Ia}(u) F^I\wedge F^a + \lambda_{ab}(u) F^a\wedge F^b.
\end{equation} We cannot in general remove the cross terms $\kappa_{Ia}(u)$
by the shift $F^I \to F^I + c_{I}^a F^a$, because $c_{I}^a$ needs to respect
the integrality of the charges. Moreover, such a shift  should accompany
nontrivial monodromy around codimension-one singularities where charged fields are massless.
Indeed, the special coordinates $a^I$, $a^D{}_I$
and  the mass parameters $m^a$ are acted upon by  such a monodromy.
Thus, their supersymmetric partners, i.e.~$F^I$,
its dual $F^D{}_I$, and $F^a$ are also mixed under the monodromy.
This mixing translates into an integral symplectic transformation law
acting on $\tau_{IJ}(u)$, $\kappa_{Ia}(u)$ and $\lambda_{ab}(u)$.
It will be quite interesting, but beyond the scope of the present paper, to determine these functions.

\section{General Features of $\cN=2$ SCFTs}\label{features}
In this section we will extract from our formalism some
generic conclusions which hold for any $\cN=2$ superconformal field theory.
Along the way, we will need to make a few assumptions,
which we will try to state explicitly.

\subsection{$2a-c$ and the dimensions of operators}\label{2a-c}
First, recall that $(2a-c)_{\infty}$ is related to the sum of the dimensions
of the Casimir operators, \eqref{group-theory-relation}.
In the two examples we saw in the last section, we also
had the formulae \eqref{2a-c:SU(n)-AD}, \eqref{2a-c:USp(2n)-AD}
relating $2a-c$ to the sum of operator dimensions. Let us repeat
the arguments we made there in general terms
to show the same relation holds for any $\cN=2$ SCFT.

Suppose a superconformal field theory is realized
at a point of the moduli space of a renormalizable gauge theory.
Let us introduce monodromy invariant coordinates
of the Coulomb branch  $u_i$
($i=1,\ldots,n$) so that the superconformal point is at $u_1=u_2=\cdots=u_n=0$.
Here $n$ is the dimension of the Coulomb branch of the original theory.
The mass of the BPS saturated soliton with electric charge $q^I$,
magnetic charge $p_I$ and the flavor charge $s^a$
 is given by the formula \begin{equation}
m(q,p,s)=|q^I a_I + p_I a^I_D + m_a s^a| \label{BPSformula}
\end{equation} where $a_I$, $a^I_D$ are the special coordinates and their duals, and
$m_a$  are the mass parameters of the theory.
At the superconformal point, two or more mutually nonlocal states simultaneously become massless,  charged under various $\UONE$ factors.
Choose $r$ so that $\UONE^r$ is the smallest gauge subgroup with respect to which all of the massless states are charged.  Then turning on a vev for the scalar component of any of the $\UONE$ factors will deform the SCFT out into the Coulomb branch of the gauge theory;  these deformations span
the $r$-dimensional Coulomb branch of the superconformal theory.

We perform a symplectic transformation of the special coordinates, including
the mass shift as in \eqref{BPSformula}, so that
$a_1,\ldots,a_r$ all go to $0$ at the superconformal point with mass dimension 1.
They are the lowest components of the $r$ vector multiplets
which couple electrically to the massless charged states at the superconformal point.
The others, $a_{r+1},\ldots,a_n$
are constant at leading order, and correspond to
vector multiplets which decouple from the SCFT.

Now the factor $A$ is
given by the square root of the Jacobian from $a_I$ to $u_i$
as in \eqref{generalconjecture}.
The scaling dimensions of the $u_i$ can be calculated from the Seiberg-Witten curves,
and we order the operators $u_i$ so that $D(u_1)\le D(u_2)\le \cdots \le D(u_n)$.
We furthermore relabel \begin{equation}
\mu_i\equiv u_i \ (i=1,\ldots,n-r),\qquad
\cO_i\equiv u_{n-r+i}\ (i=1,\ldots,r).
\end{equation}
With this preparation, we expect as in \eqref{ADleading}
that the terms which give
the leading behavior of $\det (\partial a_I/\partial  u_i)$
are the ones where the $\mu_i$'s differentiate $a_I$'s for $I=r+1,\ldots,n$
and the $\cO$'s differentiate the $a_I$'s  for $I=1,\ldots,r$.
Then we have \begin{equation}
R(A)= \sum_{i=1}^r \left(D(\cO_i)-1\right).
\end{equation} From  \eqref{generalconjecture} we have the relation
$2(2a-c)=R(A)+r/2$, so \begin{equation}
4(2a-c)=\sum_{i=1}^r (2D(\cO_i) -1).\label{2a-c-relation}
\end{equation}

So far we have defined the rank $r$ of the superconformal theory
to be the number of electric special coordinates $a_I$ which go to zero
at the superconformal points. Another natural definition of the rank
would be the number of  $u_i$'s whose scaling dimension is larger than 1,
and which thus correspond to physical operators.
In the examples we have studied, these two definitions have always agreed,
and we conjecture that this property will hold for generic $\cN=2$ SCFTs.

Before going to the next topic,
let us recall how the relation \eqref{2a-c-relation}
was originally observed  in \cite{AW}
for superconformal theories realized using the S-duality
approach of Argyres and Seiberg \cite{AS}.
The authors of \cite{AW}
started from a  finite $\cN=2$ theory with the gauge group $G$
and studied its strong coupling limit. They identified its dual  realization
as a strongly-coupled superconformal theory $S$, with flavor symmetry
group $F$ whose subgroup $G'$ is weakly gauged, together with
hypermultiplets charged under $G'$.
Therefore, they concluded that\begin{equation}
4(2a-c)_G = 4(2a-c)_{S} + 4(2a-c)_{G'},\label{awobservation1}
\end{equation} while the sets of the dimensions of Coulomb branch
operators should satisfy \begin{equation}
\{ D(u_{i,G}) \} = \{ D(u_{i,S}) \} \cup \{ D(u_{i,G'}) \}.\label{awobservation2}
\end{equation}  As in \eqref{group-theory-relation},
we have \begin{equation}
4(2a-c)_G=\sum (2 D(u_{i,G})-1)
\end{equation} and the same for $G'$.
Combined with \eqref{awobservation1}, \eqref{awobservation2}, they obtained
\begin{equation}
4(2a-c)_S=\sum (2 D(u_{i,S})-1).
\end{equation}
Our arguments generalize this observation to any $\cN=2$ superconformal
theory which can be embedded in an $\cN=2$ gauge theory.

\subsection{Bounds on the ratio $a/c$}\label{a/c}
There has been an increasing interest 
in the range of possible values of the ratio $a/c$
allowed by the causality or unitarity of the theory; see \cite{HM} and
references therein.

Our general analysis can  be used to obtain such upper and lower bounds
of the ratio $a/c$ of possible $\cN=2$ SCFTs.
First, we can combine \eqref{2a-c-relation} and the unitarity bound
$2D(\cO_i)-1\ge 1$ to show that
$4(2a-c)\ge r > 0$, i.e.
\begin{equation}
\frac12\le \frac{a}{c}.\label{lowerbound}
\end{equation}  The inequality is saturated by a free hypermultiplet.

The upper bound is harder to obtain. One way to proceed is to compare the
orders of zeros of the gravitational factors $A$ and $B$ along codimension-one
loci. Indeed it is easy to check that the order of any zero of $B$ is larger than
its order as a zero of $A$, for both types of codimension one loci we treated in Sec. \ref{codimension-one}. Then $B/A$ is holomorphic without poles on the whole moduli space
apart from possible singularities whose co-dimension is more than two.
Suppose such singularity of high codimension can be  removed so as to make
$B/A$  holomorphic throughout the moduli space.
Then we have $R(A)\le R(B)$. From \eqref{master-formula} it follows that\begin{equation}
2(2a-c)- \half r  \le 3c -\half r - \quarter h,
\end{equation} which implies the upper bound  \begin{equation}
\frac{a}{c} \le \frac{5}{4}.\label{upperbound}
\end{equation}

To make the derivation more complete,
consider the theta function
associated to the coupling functions $\tau_{IJ}(u)$, \begin{equation}
\Theta(\tau_{IJ})=\sum_{k^1,\ldots,k^n \in \bZ^n } \exp(\pi i \tau_{IJ} k^I k^J)
\end{equation} which has the modular transformation property
\begin{equation}
\Theta( (\cA\tau + \cB)(\cC\tau +\cD)^{-1} ) =\epsilon
\left[\det(\cC\tau+\cD)\right]^{1/2}\Theta(\tau),
\end{equation} where $\cA,\cB,\cC$ and $\cD$ are $n\times n$
matrices with integral entries, and $\epsilon$ is an eighth root of unity.
$\Theta$ converges as long as $\Im \tau_{IJ}$ is positive definite,
and is smooth in the limit where one of the gauge coupling goes to zero
and so $i\tau\to-\infty$.
Then $(A\Theta)^8$ is a holomorphic function on the moduli space
with neither cuts nor poles. The reason is that the monodromy of 
$\Theta$ cancels the monodromy of $A$, and $\Theta$ is smooth
at each of the codimension-one loci.

$B^8$ is also without cuts. As discussed before, it can be checked that
the order of any zero of $B^8$ is always greater than or equal to the order of
$(A\Theta)^8$ at the codimension-one loci.
Thus, $B^8/(A\Theta)^8$ is holomorphic at codimension-one loci,
and its possible singularities have codimension larger than one.
By Hartog's theorem it is guaranteed that such holomorphic functions
are in fact holomorphic throughout the moduli space,
so $B^8/(A\Theta)^8$ is holomorphic without poles nor cuts.
Hence $B$ has a higher-order zero than $A\Theta$ at the superconformal points.

To deduce $R(A)\le R(B)$ from this statement,
we need to argue that $\Theta$ is smooth at the superconformal point.
This can be checked for specific examples, like the $\SU(3)$ Argyres-Douglas point.
There, what happens is that $\tau_{IJ}$ stays constant near the superconformal
point, forcing $\Theta$  to be smooth. Another possibility is that the infrared theory
is trivially superconformal. In that case one of the gauge couplings goes to zero,
causing $\Theta$ to vanish smoothly.

We believe this property should hold in general.
When a superconformal point is at
weak coupling, the constancy of the coupling $\tau$ is one of the defining properties
of a conformal theory. In the strongly-coupled case,
we expect that the couplings $\tau_{IJ}$ will be pinned at a monodromy
invariant value, because several codimension-one loci with noncommuting
monodromy matrices will in general collide at the superconformal point \cite{AD}.
On the other extreme, when the point is free in the infrared,
some components of $\tau_{IJ}$ become infinite, which makes
$\Theta$ smoothly vanish as argued above.
Thus we expect the theta function to be smooth
at the superconformal point, which in turn implies \eqref{upperbound}.
For specific gauge theories with hyperelliptic Seiberg-Witten curves,
the argument here can be made completely rigorous
using Thomae's formula, as explained in Appendix~\ref{Thomae}.

Combining our upper and lower bounds, we have \begin{equation}
\frac12 \le \frac ac \le \frac 54.
\end{equation} The upper bound is saturated by free vector multiplets
and the lower bound  by free hypermultiplets.
It would be worthwhile to make the derivation more watertight;
it would also be interesting to consider similar bounds for superconformal
theories with fewer supersymmetries.

In
\cite{HM}  bounds on $a/c$ have been obtained based on a positive energy assumption.  For $\cN=1$ gauge theories the authors of \cite{HM} find that $\frac12 \le \frac{a}{c} \le \frac32$.  The lower inequality is saturated by theories consisting only of free hypermultiplets and the upper inequality is saturated by free vector multiplets.  Our upper bound on $a/c$ for $\cN=2$ is somewhat tighter, because $\cN=2$ vector multiplets are a combination of $\cN=1$ chiral and vector multiplets. This same bound can be obtained from the positivity of energy
in a background created by the insertion of an $\SU(2)_R$ current\footnote{
The authors would like to thank J. Maldacena for explaining this point.}.

\section{Conclusions}\label{conclusion}
The objective of this paper has been to establish
a purely field-theoretical method to calculate the central charges
of $\cN=2$ superconformal field theories. Given the usual definition
of the central charges in terms of the response of the theory to an external
gravitational field, it was natural to utilize topological twisting
to put the theory on a general curved manifold while preserving the supersymmetry.
We showed how the central charges $a$ and $c$ can be
obtained from the $\UONE_R$ anomaly of the vacuum of the topological theory, which is
related to the gravitational factors $A^\chi B^\sigma$ in the path integral measure.
The form of $A^\chi B^\sigma$  has been known for ten years, and
$a$ and $c$ can be easily obtained by evaluating the $R$-charges of these known terms.

We applied our general arguments to several specific examples,
including maximal-rank Argyres-Douglas points of pure $\SU(N)$ gauge theory,
and  $\USp(2N)$ gauge theory with $N_f=1,2,3$
fundamental hypermultiplets and one hypermultiplet in the antisymmetric
representation. The latter has a dual gravity description, from which the
central charges had been previously obtained.  Our calculation gives a purely
field-theoretical confirmation of that result.

We then discussed a few general properties of $\cN=2$ SCFTs
which come from our framework. In particular we derived the relation
of $2a-c$ to the sum of the dimensions of the operators which parameterize
the Coulomb branch. We also obtained upper and lower bounds on $a/c$.

A possible program for future research would be to try to exploit further the relationship between twisted and untwisted $\cN=2$ gauge theories, to obtain further information about these still-mysterious SCFTs.  For example, certain correlation functions might be computable by an extension of our approach.

An interesting question, which we have not addressed in this paper, is whether our results have any implications for the validity of the conjectured $a$-theorem in four dimensions.  Attempts to prove a general $a$-theorem using $a$-maximization have been complicated by the possible appearance of accidental $R$-symmetries in the infrared limit \cite{IW}.  Since our approach automatically accounts for all accidental $R$-symmetries, it is conceivable that it could be used  to prove some sort of $a$-theorem, within the class of $\cN=2$ gauge theories.

\acknowledgments
The idea that the $A^\chi B^\sigma$ factor might be related
to $a$ and $c$ was originally communicated to YT by his then-advisor
Tohru Eguchi a few years ago, and the authors are
greatly indebted to him.
They would like to thank O. Aharony, P. Argyres, M. Henningson, J. Maldacena, H. Ooguri and B. Wecht for  discussions.
They would like to thank I. V. Melnikov to point out an error which was in \eqref{error} in the version 1 of this draft.

AS gratefully acknowledges support from the Ambrose Monell Foundation and the Institute for Advanced Study.  The work of AS is also partially
supported by NSF grants PHY-0555444 and PHY-0245214.
The work of YT is in part supported by the Carl and Toby Feinberg fellowship
at the Institute for Advanced Study, and by the United States
DOE Grant DE-FG02-90ER40542.

\appendix
\section{Nonvanishing of the Jacobian}\label{ADproof}
In this Appendix for Sec.~\ref{ADac},
we prove that the contribution \begin{equation}
\left(\det \frac{\partial a_{S,i}}{\partial\cO_j} \right)
\left( \det \frac{\partial a_{L,i}}{\partial\mu_j}\right) \label{leading-in-jacobian}
\end{equation} in the expansion of the Jacobian \begin{equation}
\det\frac{\partial a_i}{\partial \tilde u_i}
\end{equation}  is nonzero close to the Argyres-Douglas point.
This fact was crucial in deriving the scaling dimension of the Jacobian, \eqref{2a-c:SU(n)-AD}.
We continue to use the notation in Sec.~\ref{ADac}.
For the $\SU(3)$ Argyres-Douglas point,
the  behavior of the special coordinates
was studied already in \cite{AD}, and the matrix $\partial a^I/\partial u_i$ was
expressed explicitly in terms of the elliptic functions in \cite{KY}.
These results suffice to show the term \eqref{leading-in-jacobian} is nonzero
for $\SU(3)$. Our aim is to extend them for general $\SU(N)$ Argyres-Douglas points.

First let us recall a fundamental fact about the Riemann surfaces.
Let $C$ be the Riemann surface of genus $g$, and let $A_i$, $B^i$
($i=1,\ldots,g$) constitute a  canonical homology basis
such that $A_i\cdot A_j=B^i\cdot B^j=0$, $A_i\cdot B^j=\delta_i^j$.
Also let $\nu^i$, $(i=1,\ldots,g)$ be a basis of
holomorphic one-forms. Then the $g\times g$ matrix $(\int_{A_i} \nu^j)$
has nonzero determinant. Indeed, if it is not invertible,
there is a nonzero holomorphic one-form $\nu$ such that $\int_{A_i}\nu \equiv 0$.
Thus we have \begin{equation}
0\ne \Im \int_C \nu \wedge \bar \nu
=\Im \sum_i\left(\int_{A_i}\nu\int_{B^i}\bar \nu
-\int_{A_i}\bar\nu\int_{B^i} \nu \right)=0,
\end{equation} which is contradictory.

Let us come back to the analysis of the Argyres-Douglas point.
We approach the superconformal point by scaling $\tilde u_i$'s by
writing them as  \begin{equation}
\tilde u_i = t^{i/(N+2)} \hat u_i
\end{equation} and taking $t\to 0$ limit keeping $\hat u_i$ fixed.
There, the genus-$(N-1)$ Seiberg-Witten curve $C$
nearly splits into two curves $C_S$ and $C_L$ of genus $r=\lfloor (N-1)/2 \rfloor$
and $r'=N-1-r$, respectively,
as depicted in Figure~\ref{adcuts}.
The former contains the short cuts $A_{S,1},\ldots,A_{S,r}$
and the latter the long cuts $A_{L,1},\ldots,A_{L,r'}$.
Let us recall  that \begin{equation}
\frac{\partial}{\partial \tilde u_i}\lambda_{SW}=\frac{x^{N-i}dx}{y}
\end{equation} are the holomorphic one-forms on $C$, and that we defined
\begin{align}
\cO_j&=\tilde u_{N-j+1} &&\text{for $j=1,\ldots,r$};\\
\mu_j&=\tilde u_{j+1} &&\text{for $j=1,\ldots,r'$}.
\end{align}
Then we have \begin{align}
\frac{\partial a_{S,i}}{\partial \cO_j}&=\int_{A_{S,i}}\frac{x^{j-1}dx}y,&&
\text{for $j=1,\ldots,r$},\label{short}\\
\frac{\partial a_{L,i}}{\partial \mu_j}&=\int_{A_{L,i}}\frac{x^{N-j-1}dx}y,&&
\text{for $j=1,\ldots,r'$}.\label{long}
\end{align}

Close to the small cuts $A_{S,i}$
it is natural to introduce the rescaled variables $\hat x$ and $\hat y$
via $x=t^{1/(N+2)} \hat x$ and $y=t^{N/(N+2)}\hat y$.
The curve $C$ of \eqref{ADcurve}  can then be approximated by
$C_S$  :
\begin{equation}
\hat y^2= 2\Lambda^N(\hat x^N+\hat u_2 \hat x^{N-2}+\cdots+\hat u_N) .
\end{equation}
The positions of the short cuts $A_{S,i}$
remain finite in the rescaled variables, and we have \begin{equation}
\frac{\partial a_{S,i}}{\partial \cO_j}=t^{(j-N)/(N+2)}
\int_{A_{S,i}}\frac{\hat x^{j-1}d\hat x}{\hat y}.
\end{equation}
Apart from the powers in $t$ as the prefactors, this
gives exactly the matrix of the pairing of the
A-cycles and the basis of holomorphic one-forms of this curve $C_S$.
From the general fact on Riemann surfaces reviewed above,
we conclude the determinant of \eqref{short} is nonzero.

The integral over  the long cuts $A_{L,i}$ can be carried out similarly.
In the $t\to 0$ limit,
 the curve $C$ can be approximated near the long cuts by
\begin{equation}
y^2= x^N (x^N+2\Lambda^N).
\end{equation} Therefore, the integral in \eqref{long} becomes \begin{equation}
\int_{A_{L,i}}\frac{x^{N-j-1}dx}y
=\int_{A_{L,i}}\frac{x^{r'-j}dx}{x^{\sigma/2}\sqrt{x^N+2\Lambda^N}}
=\int_{A_{L,i}}\frac{x^{r'-j}dx}{\tilde y}
\end{equation}where $\sigma=0$ or $1$  if $N$ is even or odd, respectively,
and we introduced the curve $C_L$ : \begin{equation}
\tilde y^2= x^\sigma(x^N+2\Lambda^N).
\end{equation}
This matrix is thus the pairing of the A-cycles and the holomorphic one-forms
of $C_L$,
which then implies that the determinant of \eqref{long} is nonzero.
This concludes the proof.

\section{Thomae's Formula and $a/c$}\label{Thomae}
The argument in Sec.~\ref{a/c}
can be made  precise
for theories with hyperelliptic Seiberg-Witten curves
by the use of Thomae's formula (see e.g. Proposition 3.6 in \cite{Fay}).
Let us first present the formula for a general
hyperelliptic curve given by $y^2=f(x)$ with a polynomial
$f(x)$ of degree $2n$. We split $2n$ zeroes of $f(x)$ to two sets,
$e_a$ and $e'_a$, $(a=1,\ldots,n)$. We then choose $I$-th A-cycles of the curve
to be the path encircling $e_I$ and $e'_I$.
We take $\tau_{IJ}$ to be the period matrix in this basis.
Then the formula states
\begin{equation}
\prod_{a>b}(e_a-e_b)^2\prod_{a>b}(e'_a-e'_b)^2
=k\left(\det M_{ij}\right)^{-4} \Theta[\substack{\delta \\\delta '}](\tau_{IJ})^8.
\label{thomae-formula}
\end{equation}
Here, $M_{ij}$ is the matrix \begin{equation}
M_{ij}=\int_{A_i} \frac{x^{j-1} dx}{y}, \qquad (i,j=1,\ldots,n)
\end{equation}  which pairs  the A-cycles with the holomorphic differentials,
\begin{equation}
\Theta[\substack{\delta \\\delta '}](\tau_{IJ})
=\sum_{k^1,\ldots,k^n \in \bZ^n }
\exp\left[\pi i \tau_{IJ} (k^I+\delta^I)(k^J+\delta^J)+2 \pi i (k^I+\delta^I) \delta'_I\right]
\end{equation} is the theta function with a particular half-integer characteristic
$[\substack{\delta\\ \delta'}]$ determined by the choice of the A-cycles,
and $k$ is a nonzero constant.

We can rewrite this relation using the gravitational measure factors $A(u)$ and $B(u)$.
First, we have \begin{equation}
M_{ij}=\int_{A_i} \frac{\partial}{\partial u_j}\lambda_{SW} =
\frac{\partial a_i}{\partial u_j}
\end{equation} from the defining property of the Seiberg-Witten differential.
Therefore $\det M_{ij}= A(u)^{-2}$.
Second, recall that the Seiberg-Witten curve is given by \begin{equation}
y^2=(P(x)+\Lambda^N)(P(x)-\Lambda^N)
\end{equation}
for pure $\SU(N)$ gauge theory,
and the conventional choice of the $A$-cycles corresponds to the
assignment of $e_{a}$, $e'_a$ ($a=1,\ldots,N$) to the zeroes
of $P(x)+\Lambda^N$ and $P(x)-\Lambda^N$, respectively.
The discriminant $\Delta$ is, by definition,
\begin{equation}
\Delta=\prod_{a>b} (e_a-e_b)^2 \prod_{a>b} (e'_a-e'_b)^2 \prod_{a,b}(e_a-e'_b)^2.
\end{equation}
For this particular Seiberg-Witten curve, $e_a-e'_b$ never vanishes because
\begin{equation}
P(e_a)-\Lambda^N=(P(e_a)+\Lambda^N)-2\Lambda^N=-2\Lambda^N
\end{equation} can never vanish.
Thus, $\Delta$ is equal to the left hand side of
Thomae's formula \eqref{thomae-formula} up to a constant factor.
Therefore, Thomae's formula implies in this case  \begin{equation}
B^8(u)=k' A(u)^8 \Theta[\substack{\delta\\ \delta'}](\tau_{IJ}(u))^8
\end{equation}  with another nonzero constant $k'$.
This  means then that the order of zero of $B(u)$ is
always higher than or equal to that of $A(u)$,
because theta functions can only have zeroes but not poles.
Therefore we conclude $R(A)\le R(B)$. This implies, as in Sec.~\ref{a/c},
that $a/c$ is bounded above by $5/4$.

\end{document}